\documentclass[11pt]{article}
\usepackage{hyperref}
\pdfoutput=1
\begin{document}
\title{Fast and Global Measurement \\ of Free Surface Deformations}
\author{Cobelli P., Lagubeau G., \\ Chekroun M., Przadka A., \\ Maurel A., Pagneux V. and Petitjeans P. \\ \\
 \vspace{6pt}  \\ Laboratoire PMMH, UMR CNRS 7636 \\ ESPCI - ParisTech \\ Institut Langevin - LOA \\ ESPCI - ParisTech\\ UMR CNRS 7587 \\ 
 \\ LAUM \\ UniversitŽ du Maine -  UMR CNRS 6613} 
\maketitle
\begin{abstract}
In this fluid dynamics video, we present an overview of the most important results recently obtained in our group using an 
 optical profilometric
technique that allows for single-shot global
measurement of free-surface deformations. This system
consists of a high-resolution system composed of a videoprojector
and a digital camera. A fringe pattern of known
characteristics is projected onto the free surface and its
image is registered by the camera. The deformed fringe
pattern arising from the surface deformations is later
compared to the undeformed (reference) one, leading to a
phase map from which the free surface can be reconstructed.

Two examples are developed: (a) the study of water wave trapped modes and (b)
the impact of a drop onto a thin layer of the same liquid. 
\end{abstract}
\section{Description of what is shown in the fluid dynamics video}

The video first presents the principle of the Fourier Transform Profilometry (FTP) technique due to Takeda \& Mutoh (1983). Its applicability to the case of free surfaces is illustrated by the time measurement of a shaken liquid.

The application of this technique to two subjects of current interest in research are shown. 

The first one corresponds to the occurrence of trapped modes around and obstacle in a water waveguide (Cobelli et al. 2010). The relevant parameters here are the aspect ratio between the cylinder diameter $a = 5$~cm and the guide width $d = 10$~cm, ($a/d = 0.5$), and the trapped mode frequency 2.5~Hz. A decomposition in harmonics of the driving frequency is shown, as well as an spatial separation in symmetrical and antisymmetrical components (with respect to the longitudinal axis of the waveguide). 

The second subject concerns the flower pattern observed a few miliseconds after the impact of a drop onto a thin layer of liquid (Lagubeau et al., PRL 2010). The case shown in the video corresponds to the impact of a water drop of 4.1~mm in diameter onto a water layer of 3.1~mm thickness. The falling height is 30~cm, which corresponds to a Weber number $We = 110$. Inmediately after the impact a rim is produced that propagates outwards radially. This rim becomes unstable and approximately 30~ms after, when the rim relaxes, a wave pattern is emitted whose symmetry is related to the wavelength of the initial instability. The video shows this wave pattern as well as the jet consequently emitted at the point of impact.  

\section{References}

\noindent Takeda M, Mutoh K  Fourier transform profilometry for the
automatic measurement of 3-D object shapes. Appl Opt 22 (1983). \\

\noindent Cobelli P., Maurel A., Pagneux V., Petitjeans P. 
Global measurement of water waves by Fourier transform profilometry
Experiments in Fluids, Vol. 46, Issue 6 (2009). \\ 

\noindent Cobelli P., Pagneux V., Maurel A., Petitjeans P. 
Experimental observation of trapped modes in a water wave channel
Europhysics Letters, Vol. 88:2, 20006 (2009). \\ 

\noindent Cobelli P., Pagneux V., Maurel A., Petitjeans P. 
Experimental study on water-wave trapped modes, JFM (2010). \\

\noindent Lagubeau G., Fontelos M., Josserand C., Maurel A., Pagneux V. and Petitjeans P.
Flower patterns in drop impact, PRL (2010). \\

%
%
\end{document}